\documentclass{ifacconf}

\usepackage{amssymb,amsmath}
\usepackage{booktabs}
\usepackage{graphicx}      
\usepackage{natbib}        
\usepackage{mathrsfs}
\begin{document}
\begin{frontmatter}

\title{Univariate ReLU neural network and its application in nonlinear system identification\thanksref{footnoteinfo}} 

\thanks[footnoteinfo]{This work is jointly supported by the National Natural Science Foundation of China (U1813224), and Science and Technology Innovation Committee of Shenzhen Municipality (JCYJ2017-0811- 155131785).}

\author[First]{Xinglong Liang} 
\author[Second]{Jun Xu} 

\address[First]{Harbin Institute of Technology, Shenzhen (e-mail: liangxl814@163.com)}
\address[Second]{Harbin Institute of Technology, Shenzhen (e-mail: xujunqgy@hit.edu.cn)}

\begin{abstract}                
ReLU (rectified linear units) neural network has received significant attention since its emergence. In this paper, a univariate ReLU (UReLU) neural network is proposed to both modelling the nonlinear dynamic system and revealing insights about the system. Specifically, the neural network consists of neurons with linear and UReLU activation functions, and the UReLU functions are defined as the ReLU functions respect to each dimension. The UReLU neural network is a single hidden layer neural network, and the structure is relatively simple. The initialization of the neural network employs the decoupling method, which provides a good initialization and some insight into the nonlinear system. Compared with normal ReLU neural network, the number of parameters of UReLU network is less, but it still provide a good approximation of the nonlinear dynamic system. The performance of the UReLU neural network is shown through a Hysteretic benchmark system: the Bouc-Wen system. Simulation results verify the effectiveness of the proposed method.

 
\end{abstract}


\end{frontmatter}

\section{Introduction}
System identification deals with the problem of building mathmatical models of dynamical systems based on observed data from the system (\cite{lennart1999system}), which can be applied in industrial processes, economic and financial systems, biology and life sciences, medicine, social systems, and so on (\cite{billings2013nonlinear}). Since the theory of linear time-invariant (LTI) systems has been extensively studied over
decades (\cite{lathi1998signal}), a huge amount of effective system identification methods have been developed for linear systems (\cite{nelles2013nonlinear, billings2013nonlinear}).  However most industrial systems are nonlinear systems, for which accurate descriptions can not be built by just using the identification methods developed for linear systems. Hence 
the nonlinear system identification is currently a field of active research (\cite{westwick2018using};\cite{schoukens2016linear};\cite{abdalmoaty2018application};\cite{dreesen2017modeling}).

As an effective method, neural networks based method has played important roles in different fields, which is due to that they are conceptually simple and easy to train and use (\cite{billings2013nonlinear}). 
 For example, \cite{chen1992neural} has employed 3 kinds of neural networks in the identification of nonlinear discrete-time dynamic systems. Besides, \cite{billings2005new} has proposed a new class of wavelet networks for nonlinear system identification which exploits the attractive features of both multiscale wavelet decompositions and the approximation capability of traditional neural networks. In \cite{ganjefar2015single}, a single hidden layer fuzzy recurrent wavelet neural network has been developed for function approximation and dynamic system identification, which could achieve a higher accuracy with a lower number of parameters compared with other models. Then \cite{yao2018identification} has proposed an improved identification method based on sinusoidal echo state network  to identify a class of periodic discrete-time nonlinear dynamic systems with or without noise. 
 
 In spite of the excellent performance of neural networks, they still receive criticism as they are basically black-box models, making it difficult to draw any insights from the identified model. Moreover, the parameters of the neural network may be too much, and its training requires sophisticated skills and tunings.

In neural networks, the rectified linear units (ReLU) is an effective kind of activation functions, and the effectiveness is more clearly shown  in deep neural networks. Compared with other activation functions like Sigmoid, ReLU effectively solved the problem of vanishing gradient (\cite{hochreiter1998vanishing}),  which is caused by the saturation of activation functions. Saturation implies that the derivative of the activation converges to zero as the input goes to both $+\infty$ and $-\infty$ (\cite{neyshabur2016path}). Then, as the derivative of the  ReLU activation function is $1 $ or $0$, using ReLU as the activation function effectively alleviates the problem of gradient vanishing and it also leads to the sparsity of the network, which is useful in deep network and improves the accuracy of learning.  

In this paper, we propose a univariate ReLU (UReLU) neural network based on the UReLU activation function, which is ReLU function with respect to each dimension. The initialization of the UReLU network can be fulfilled by a decoupling method, which was based on tensor decomposition and proposed by \cite{westwick2018using} and \cite{dreesen2018decoupling}. After initialization, the parameters of the UReLU neural network can be estimated by the variable projection method.

The rest of the paper is organized as follows. Section 2 gives a detailed description of the UReLU function, and then the structure as well as the training of the UReLU neural network are provided in Section 3 and 4, respectively. Section 5 describes the interpretability of the UReLU neural network, which will facilitate the subsequent control or optimization after nonlinear system identification. Simulation studies are shown in Section 6. Finally, Section 7 ends with conclusions and future work.

\subsection{Notation}
Lower and upper case letters in a regular type-face, $a$, $A$, will refer to scalars, bold faced lower case letters, $\mathbf a$ refer to vectors, matrices are indicated by bold faced upper case letters, $\mathbf A$, and bold faced calligraphic script will be used for tensors, $\mathbf{\mathcal{A}}$.
The notation $x_i$ denotes the $i$-th component in $x$, while the notation $A_{ij}$ denotes the element in the $i$-th row, $j$-th column of $A$, and $A(:,i)$ denotes the $i$-th column in the matrix $\mathbf A$. The superscript ``$T$" denotes the transpose. The real scalar set is denoted by $\mathbb{R}$ while the real vector set in $n$ dimension is written as $\mathbb{R}^n$. 

\section{Univariate ReLU function}

As mentioned earlier, the rectified linear units (ReLU) $\max(0,x)$ as an activation function has been used in neural networks widely. One advantage of ReLU is its non-saturating nonlinearity. In terms of training time with gradient descent, these non-saturating nonlinearity are much faster than the saturating nonlinearity like sigmoid function (\cite{krizhevsky2012imagenet}).  Another nice property is that compared with sigmoid function, the derivative of ReLU can be implemented by simply thresholding a matrix of activations at zero.

Specifically, the output of the ReLU can be written as 
\[\max\{0,\mathbf w^T\mathbf x+ b\},\]
where $\mathbf{x}=[x_1, \ldots, x_n]^T \in \mathbb{R}^n$, $\mathbf w \in \mathbb{R}^n$ is the weight matrix and $b$ is the bias. 

In this paper, the univariate ReLU (UReLU) is introduced as the activation function of the neural network, which can be simply expressed as
\[
\max\{0,x_i-\beta_{i1}\},\max\{0,x_i-\beta_{i2}\},...,\max\{0,x_i-\beta_{i,q_i}\}
\]
where $i \in \{1, \ldots, n\}$, and the bias $\beta_{ij} (j=1, \ldots, q_i)$ is chosen based on the training data and satisfies
\[
\beta_{i1}<\beta_{i2}<...<\beta_{i, q_i}.
\] 
Specifically, assume the training data is 
\[
(\mathbf{x}(k), y(k))_{k=1}^N,
\]
we can choose the bias $\beta_{i1}, \ldots, \beta_{i, q_i}$ according to the distribution of $x_i(k), i=1, \ldots, n, k=1, \ldots, N$. 
 For example, for the dimension $i$, we can choose $\beta_{i1}, \ldots, \beta_{i, q_i}$ to be the $q_i$-quantiles of $x_i(k), k=1, \ldots, N$.

Correspondingly, the derivative of the UReLU can be expressed as
\[
(\max\{0,x_i-\beta_{ij}\})^{\prime}=\left\{
\begin{array}{ll}
0 &~~ \mbox{if}~ x_i-\beta_{ij}\leq 0\\
1 &~~ \mbox{if}~ x_i-\beta_{ij} > 0.
\end{array}\right.
\]
Compared with the ReLU, the UReLU focuses on the single variable and the bias parameters are determined according to the data distribution and need not to be trained during the network training. Yet the UReLU retains the nice property, i.e., its derivative also can be implemented by simply thresholding a matrix of activations at zero.

Next we will explain in detail of the role of the UReLU activation function in the UReLU neural network.

\section{Structure of the UReLU neural network}
\begin{figure}[h]
	\begin{center}
			\includegraphics[width=0.8\columnwidth]{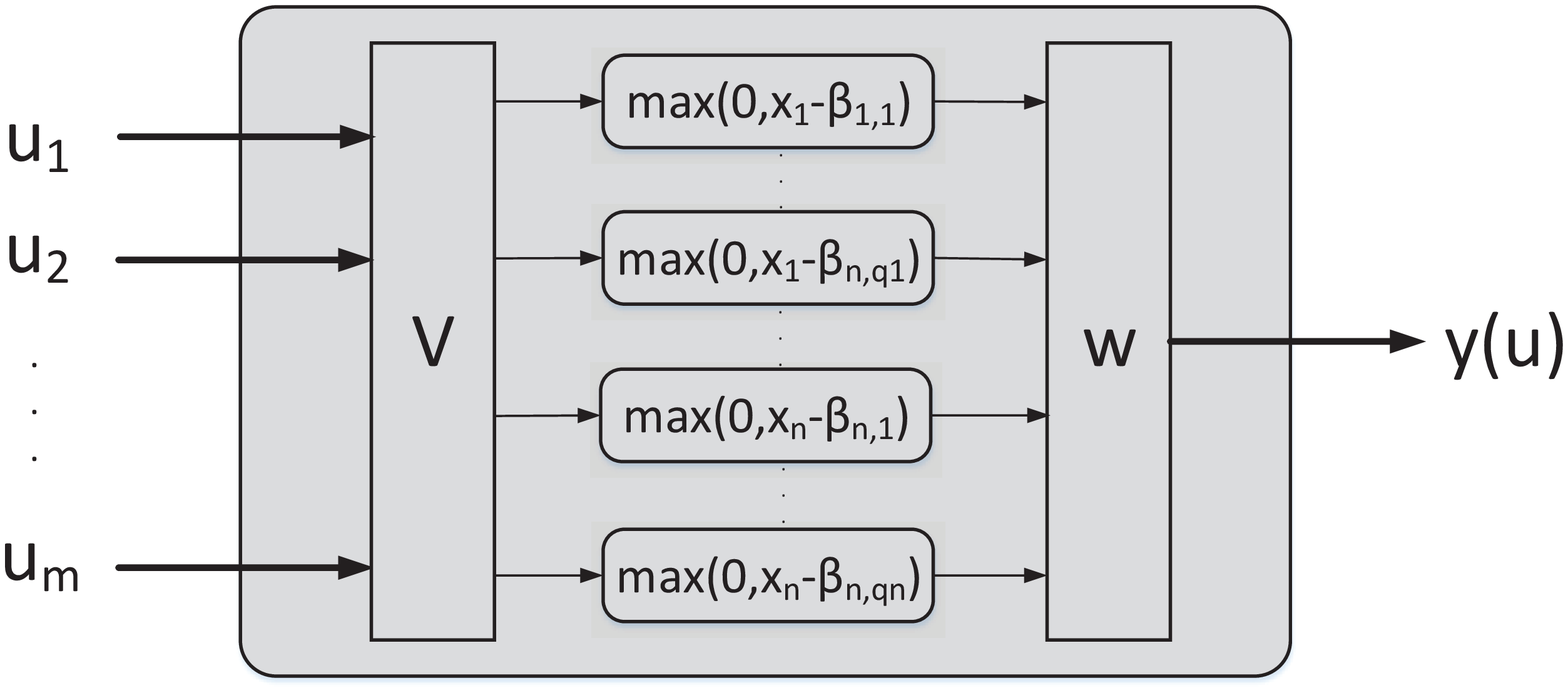}    
		\caption{Structure of the UReLU neural network} 
		\label{fig:Structure}
	\end{center}
\end{figure}

The structure of the UReLU neural network is shown in Fig. \ref{fig:Structure}, in which the input vector $\mathbf u=[u_1, \ldots, u_m]\in \mathbb{R}^m$ and the output is $y(\mathbf u) \in \mathbb{R}$. A linear transformation $\mathbf V$ is introduced to transform the input vector $\mathbf u$ to the intermediate vector $\mathbf x \in \mathbb{R}^n$, which is then sent to the UReLU neurons. Finally, the output is derived as the weighted sum of the UReLU neurons. Generally speaking, the dimension of the intermediate vector $\mathbf x$ is lower than that of the input vector $\mathbf u$, i.e., $n<m$. The UReLU neural network can be seen as a decoupled structure, i.e., the regular ReLU nonlinearity has been decoupled into several UReLU nonlinearities.

The following describes the UReLU neural network with respect to 3 parts, the connection between the input and the intermediate vector, the generation of the UReLU nonlinearities, and the connection to the output.


\subsection{Connection between the input and the intermediate vector}
The intermediate vector $\mathbf x \in \mathbb{R}^n$ is obtained through a linear transformation of the input vector $\mathbf u \in \mathbb{R}^m$, which is fulfilled through a linear transformation matrix $\mathbf V$. Assume the linear transformation matrix $\mathbf V$ can be expressed as 
 \begin{equation}
\mathbf V=\left[\begin{array}{ccc}
v_{11}&\cdots&v_{1n}\\
\vdots&\ddots&\vdots\\
v_{m1}&\cdots&v_{mn}
\end{array}\right],
\end{equation} 
then we have
\[
\mathbf x=\mathbf V^T \cdot \mathbf u.
\]

After the linear transformation, the dimension of the intermediate vector is less than that of the input vector, i.e., $n<m$.  

Assume there are $N$ samples, define the input data matrix as
\[
\mathbf U=\left[\begin{array}{ccc}
u_1(1)&\cdots&u_m(1)\\
\vdots&\ddots&\vdots\\
u_1(N)&\cdots&u_m(N)
\end{array}\right],
\]
and similarly define the intermediate data matrix as
\[
\mathbf X=\left[\begin{array}{ccc}
x_1(1)&\cdots&x_n(1)\\
\vdots&\ddots&\vdots\\
x_1(N)&\cdots&x_n(N)
\end{array}\right],
\]
then we have
 \begin{equation}
 \label{UV}
\mathbf X=\mathbf U\mathbf V
\end{equation}

\subsection{Generation of the UReLU nonlinearities}

Based on the intermediate data matrix $\mathbf X$, the UReLU nonlinearities can be represented as follows,
\begin{equation}\label{BV}
\begin{array}{rl}
\mathbf B=&[\max\{0,\mathbf x_1-\beta_{1,1}\},\ldots, \max\{0,\mathbf x_1-\beta_{1,q_1}\}\\
{}&\ldots, \max\{0,\mathbf x_n-\beta_{n,1}\}, \ldots, \max\{0,\mathbf x_n-\beta_{n,q_n}\}],
\end{array}
\end{equation} 
in which
\begin{equation}\label{xi}
\mathbf x_i=[x_i(1), \ldots, x_i(N)]^T \in \mathbb{R}^N.
\end{equation}
Hence we have
\[
\mathbf X=[\mathbf x_1, \ldots, \mathbf x_n].
\]
According to \eqref{UV}, we then have
\begin{equation}\label{xiuv}
\mathbf x_i=\mathbf U \cdot [v_{1i}, \ldots, v_{mi}]^T.
\end{equation}

As is mentioned before, the bias $\beta_{ij}, i=1, \ldots, n, j=1, \ldots, q_i$ can be determined according to the sampled data distribution. For balanced data, we set $q_i=q, \forall i \in \{1, \ldots, n\}$ and employ the following simple method to obtain the bias,
 \begin{equation}
 \label{beta}
 \beta_{ij}=[\max(\mathbf x_i)-\min(\mathbf x_i)]\cdot s_j+\min(\mathbf x_i),
  \end{equation} 
  in which
 \begin{equation}
 \label{S}
 \mathbf s=[0,1/q,...,(q-1)/q].
 \end{equation} 
 In the training of the UReLU neural network, $q$ is preset and not changed, while $\beta_{ij}$ fluctuates as $\mathbf V$ changes. Hence the data matrix $\mathbf B$ is basically dependent on $\mathbf V$, i.e., $\mathbf B(\mathbf V)$.

\subsection{Connection to the output}

In this paper, we only consider single output and denote the output data vector as
\[
\mathbf y=[y(1), \ldots, y(N)]^T.
\]

The output of the UReLU neural network can be written as:
\begin{equation}
\label{objfun}
 \hat{\mathbf y}=[\mathbf 1, \mathbf B(\mathbf V)]\cdot \mathbf w
\end{equation}
where $\mathbf{1}\in \mathbb{R}^N$ is a vector with all entries being 1, $\mathbf w=\left[w_0, w_1, \ldots, w_{M}\right]^T$ is the weight vector, $w_0$ is for the constant neuron, $M=nq$ denotes the number of UReLU neurons. $\hat{\mathbf y}$ is the predicted output, which is dependent on $\mathbf V, \mathbf w$ and the input $\mathbf u$. 



\section{Training of the UReLU neural network}

As is mentioned before, the predicted output is a function of the linear transformation matrix $\mathbf V$, the weights vector $\mathbf w$ and the input $\mathbf u$. Hence the training of the UReLU neural network deals with the problem of finding the optimal $\mathbf V$ and $\mathbf w$.
Here the gradient-based optimization method is employed, which includes the procedures of parameter initialization and optimization.

\subsection{Parameter initialization} 
A good initialization parameter is vital for the gradient based method, which will make the convergence faster and the result more stable. The initialization of the parameters $\mathbf V$ and $\mathbf w$ is fulfilled through a tensor decomposition method proposed in \cite{dreesen2018decoupling}.

Specifically, the initialization of the linear transformation matrix $\mathbf V$ consists of the following 3 steps:
\begin{description}
\item [1)] An NARX polynomial model is established based on the input and output of nonlinear system, i.e.,
\begin{equation}
\label{equ:ploy}
\begin{aligned} 
\hat y(\mathbf u)=y_0+\sum_{i_1=1}^{m}a_{i_1}u_{i_1}+\sum_{i_1=1}^{m}\sum_{i_2=i_1}^{m}a_{i_1,i_2}u_{i_1}u_{i_2}\\+...+\sum_{i_1=1}^{m}...\sum_{i_n=i_{n-1}}^{m}a_{i_1,...,i_n}u_{i_1}...u_{i_n}
\end{aligned}
\end{equation}
It is noted that we can use the forward regression with orthogonal least squares (FROLS) developed by \cite{billings1988identification} to reduce the number of polynomial terms.
\item [2)] Obtain the Hessian of \eqref{equ:ploy}. The Hessian matrix is
\begin{equation}
\mathbf H_y(\mathbf u)=\frac{\partial^{2}\hat y(\mathbf u)}{\partial{\mathbf u}^{2}}
\end{equation}
And then Hessian is evaluated at $N$ different operating points, the results can be stacked into a 3 dimensional tensor $\mathbf{\mathcal{H}}$.
\item [3)] The 3 dimensional tensor $\mathbf{\mathcal{H}}$ can be written as 
 \begin{equation}
\label{equ:H3}
\mathbf{\mathcal{H}}=\sum_{i=1}^{r}\mathbf v_\ell\otimes \mathbf v_\ell\otimes\mathbf w_\ell
\end{equation}
It can also be abbreviated as
\begin{equation}
\label{equ:H4}
\mathbf{\mathcal{H}}=[\![\mathbf V,\mathbf V,\mathbf W]\!]
\end{equation}
\end{description}
This is the Canonical Polyadic Decomposition (CPD) (\cite{kolda2009tensor}) of the tensor $\mathbf{\mathcal{H}}$. The CPD can be implemented by the tensorlab toolbox (\cite{vervliet2016tensorlab}). we only use the obtained $\mathbf V$ to initialize the linear transformation matrix. After the initialization of $\mathbf V$, the weights vector $\mathbf w$ can be obtained by least square methods, which will be explained in the next section. Actually by using this initialization, the Hessian information can provides some insight into the nonlinear system, thus facilitate the interpretability(\cite{dreesen2018decoupling}).

\begin{figure}[h]
	\begin{center}
		\includegraphics[width=0.8\columnwidth]{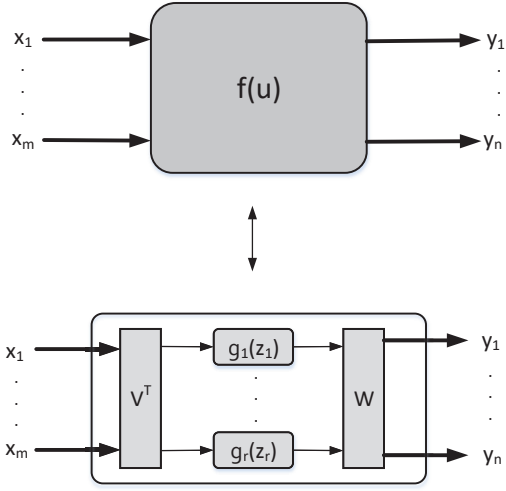}    
		\caption{A typical decoupling structure.} 
		\label{fig:decoupling}
	\end{center}
\end{figure}

The reason of using this kind of parameter initialization strategy is that the UReLU structure is similar to the decoupling structure proposed in \cite{dreesen2018decoupling}, which is shown in Fig. \ref{fig:decoupling}, in which
\begin{equation}
\mathbf f(\mathbf x)=\mathbf W \mathbf g(\mathbf V^T\mathbf x)
\end{equation}
where $\mathbf W$ and $\mathbf V$ are transformation matrices, the vector function $\mathbf g(z)=[g_1(z_1),...,g_1(z_r)]$ is composed of univariate functions $g_i(z_i)$ in its $r$ components. 



\subsection{Optimization}
Given the inputs and outputs of a nonlinear system at $N$ sampling points, our objective is to minimize the following criterion
\begin{equation}
\label{SLS}
||r(\mathbf V,\mathbf w)||_{2}^{2}=||\mathbf y-\mathbf{\hat{y}}(\mathbf V,\mathbf w)||_{2}^{2}
\end{equation}
with respect to $\mathbf V$ and $\mathbf w$.

It should be noted that $\mathbf{\hat{y}}(\mathbf V,\mathbf w)=\mathbf {B(V)}\mathbf w$ is a nonlinear function of $\mathbf V$ and a linear function of $\mathbf w$. If the matrix $\mathbf V$ is fixed, $\mathbf w$ can be easily obtained by solving the linear least squares problem:
\begin{equation}
\label{Moore}
\mathbf w=\mathbf {B(V)}^\dagger\mathbf y
\end{equation}
where $\mathbf {B(V)}^\dagger$ is the Moore-Penrose generalized inverse of $\mathbf {B(V)}$. Substituting \eqref{Moore} into \eqref{SLS} and we obtain:
\begin{equation}
\label{vapro}
\min\limits_{\mathbf V}\frac{1}{2}||r(\mathbf V,\mathbf w)||_{2}^{2}=\min\limits_{\mathbf V}\frac{1}{2}||\mathbf y-{\mathbf {B(V)}}{\mathbf {B(V)}}^\dagger{\mathbf y}||_{2}^{2}
\end{equation}
This is the Variable Projection functional (\cite{golub2003separable}). 
In this case, the linear parameter $\mathbf w$ depends on the nonlinear parameters $\mathbf V$, hence we only need to focus on optimizing $\mathbf V$.

To solve the problem \eqref{vapro}, the most reliable nonlinear least-squares algorithms require the Jacobian matrix 
\begin{equation}\label{rv_partial}
\frac{\partial r} {\partial \mathbf V}.
\end{equation}
 \cite{Lawton1971Elimination} used finite differences to obtain the approximated derivatives for \eqref{vapro} and satisfactory results were obtained. In 1973, Golub and Pereyra showed how the Jacobian (\ref{rv_partial}) could be computed exactly from the derivative
 \begin{equation}\label{bv_partial}
 \frac{\partial \mathbf B(\mathbf V)}{\partial \mathbf V}.
 \end{equation}
  This was an important step for efficiency and reliability of the method.

Thanks to the structure of the UReLU neural network, the derivative (\ref{bv_partial}) is easy to be obtained.

For the element $\max\{x_i(k)-\beta_{ij}\}$ in $\mathbf B$, we will illustrate how to calculate the derivative
\[
\frac{\partial \max\{x_i(k)-\beta_{ij}\}}{\partial \mathbf V}.
\]

For the variable $v_{st}, s=1, \ldots, m, t = 1, \ldots, r$ in $\mathbf V$, we have
\begin{equation}
\label{dd}
\dfrac{\partial \max\{0, x_i(k)-\beta_{ij}\}}{v_{st}}=\left\{\begin{array}{ll}
\dfrac{\partial  x_i(k)}{v_{st}}-\dfrac{\partial \beta_{ij}}{v_{st}}& x_i(k)-\beta_{ij} > 0\\
0&x_i(k)-\beta_{ij}  \leq 0
\end{array}\right.
\end{equation}
From \eqref{xiuv}, we have
\begin{equation}
x_i(k)=u_1(k)v_{1i}+\ldots+u_m(k)v_{mi}
\end{equation}
hence
\begin{equation}
\label{dx}
\dfrac{\partial x_i(k)}{v_{st}}=\left\{\begin{array}{cc}
u_s(k)& t=i\\
0& t \neq i.
\end{array}\right.
\end{equation}
From \eqref{beta}  and \eqref{S}, we have
\begin{equation}
\begin{aligned} 
\label{dbeta}
\dfrac{\partial \beta_{ij}}{\partial v_{st}}=&s_j \cdot \frac{\partial \left[\max(\mathbf x_i)-\min(\mathbf x_i)\right]}{\partial v_{st}}-\frac{\partial \min(\mathbf x_i)}{\partial v_{st}}.
\end{aligned}
\end{equation}
Assume 
\[
x_i(k_{\min, i})=\min(\mathbf x_i), x_i(k_{\max,i})=\max(\mathbf x_i),
\]
the derivative \eqref{dbeta} can be further expressed as
\begin{equation}\label{dbeta_vst}
\begin{aligned}
\frac{\partial \beta_{ij}}{\partial v_{st}}=\left\{\begin{array}{cc}
s_j[u_s(k_{\max,i})-u_s(k_{\min, i})]-u_s(k_{\min, i})& t=i\\
0&t \neq i.
\end{array}\right.
\end{aligned}
\end{equation}

From \eqref{dx} and \eqref{dbeta_vst}, \eqref{bv_partial} can be easily obtained and the Jacobian \eqref{rv_partial} can also be calculated using methods  based on matrices and tensors as described in \cite{golub2003separable}. Finally, the problem can be solved by quasi-Newton method or Levenberg-Marquardt method, in which the number of iterations can be tuned to prevent overfitting.


\section{Interpretability of the UReLU neural network}
As is mentioned before, although most nonlinear models can capture the nonlinear phenomenon very well, the number of parameters used is large and at the same time, the model interpretability is lost.  In this section, the model interpretability is illustrated through 2 aspects: dimensionality reduction and the easy-get piecewise linear (PWL) relationship.

\subsection{Dimensionality reduction}
The transformation $\mathbf{X= U V}$ maps a data vector $\mathbf u_i$ from an original space of $m$ variables to a new space of $n$ variables. As mentioned before, we require that $n<m$. After the optimization process in the training of the UReLU neural network, the new generated data matrix $\mathbf X$ is always uncorrelated, which will be shown clearly in the simulation study. Hence, through this process, the dimensionality of the problem can be greatly reduced, which makes the subsequent training or identification problem easier.

This dimensionality reduction procedure is similar to the principle analysis (PCA), in which as much of the variance in the dataset as possible is retained. The major difference between our linear transformation and the PCA transformation is that what we focus on is the training efficiency, i.e., the variable selection procedure in this paper is a supervised process. In special, the initial linear transformation is obtained through the stacked Hession of the nonlinear dynamic system, which reflects the insights about the system to some extent. In the simulation study, we can see that this supervised selection gives a satisfactory precision and at the same time, provides a better understanding of the nonlinear dynamic system.

\subsection{Linear relationships on subregions}
The system input $\mathbf U$ passes through the dimensionality reduction module and is sent to the UReLU module. And from the expression of the UReLU functions, we can know the domain partition of the input $\mathbf U$, and in each subregion, the predicted output is affine. For the UReLU neural network, the subregions and linear relationships defined on the subregions are clear. In special, assume $\beta_{ij}$ are chosen according to 
\eqref{beta}  and \eqref{S}, there are totally $(q-1)^n$ subregions, which is the Cartesian product of the sets
\[
\begin{array}{l}
\{\beta_{11}\leq x_1 \leq \beta_{12}, \ldots, \beta_{1, q-1}\leq x_1 \leq \beta_{1q}\}\\
\vdots\\
\{\beta_{n1}\leq x_n \leq \beta_{n2}, \ldots, \beta_{n, q-1}\leq x_n \leq \beta_{nq}\},
\end{array}
\]
i.e., we have the subregions $\Gamma_{k_1\cdots k_n}$ in the $\mathbf x$ space with $k_i \in \{1, \ldots, q-1\}$ and 
\[
\Gamma_{k_1 \cdots k_n}=\left\{ x_1\in [\beta_{1, k_1}, \beta_{1, k_1+1}], \cdots, x_n \in [\beta_{n, k_n}, \beta_{n, k_n+1}]\right\}.
\]

For any $\mathbf{x} \in \Gamma_{k_1 \cdots k_n}$, we have
\begin{equation}\label{linear_function}
\hat{y}=w_0+\sum\limits_{i=1}^{k_1}w_i (x_1-\beta_{1i})+\ldots+\sum\limits_{i=1}^{k_n}w_{c_{n-1}+i}(x_n-\beta_{ni}),
\end{equation}
in which $c_{s}=sq, s=1, \ldots, n$ and $c_n=M$.
 
 The subregions $\Gamma_{k_1\cdots k_n}$ relates to the polyhedral in the $\mathbf{u}$ space through the linear transformation $\mathbf V$, i.e.,
 \[
 \mathbf V^T \mathbf u \in \Gamma_{k_1 \cdots k_n},
 \]
 besides, the linear function \eqref{linear_function} can be expressed with respect to $\mathbf{u}$ according to the linear transformation matrix $\mathbf V$.
 Therefore, the subregions as well as the linear functions in the subregions can be easily obtained, which will facilitate greatly the control and optimization of the nonlinear system after system identification.

\section{Benchmark Result}
In this section, we shall apply the UReLU neural network to
nonlinear system identification. The Nonlinear AutoRegressive eXogenous Input (NARX) model is widely used for the estimation (\cite{Billings2013}), i.e., the output of the NARX model can be written as
\begin{equation}
\begin{array}{cc}
y(t)&=g(u(t), u(t-1), \ldots, u(t-n_u),\\
{}&\quad\quad y(t-1), \ldots, y(t-n_y))+e(t)
\end{array}
\end{equation}
where $t$ is the discrete time index, $u(t)$ and $y(t)$ are the input and output, respectively, and $e(t)$ is the error term. The regression vector is defined as
\[
\varphi(t)=[u(t), u(t-1), \ldots, u(t-n_u), y(t-1), \ldots, y(t-n_y)]^T.
\]
The regressors pass through the UReLU neural network, which corresponds to the nonlinear function $g$ and the identification goal is to find the parameters of the UReLU neural network such that the error term is minimized.

To test the model on the validation set, only the input is used to
generate the simulated output, that is
\begin{equation*}
\begin{array}{cc}
y_s(t)&=g(u(t), u(t-1), \ldots, u(t-n_u),\\
{}&\quad\quad y_s(t-1), \ldots, y_s(t-n_y)).
\end{array}
\end{equation*}

\subsection{Bouc-Wen Benchmark Model}
The Bouc-Wen model has been intensively exploited during the last decades to represent hysteretic effects in mechanical engineering, especially in the case of random vibrations \cite{noel2016hysteretic}. The Bouc-Wen oscillator with a single mass, is governed by the Newtons law of dynamics written in the form of
\begin{equation}\label{1}
m_L+k_Ly+c_L\dot{y}+z(y,\dot{y})=u(t),
\end{equation}
where $m_L$ is the mass constant, $k_L$ and $c_L$ are the linear stiffness and viscous damping coefficients, respectively. Besides, $y$ is the displacement, and an over-dot indicates a derivative with respect to the time variable $t$.  The input is the external force $u$, and the nonlinear term $z(y,\dot{y})$  encodes the hysteretic memory of the system and represents the hysteretic force, i.e., 
\begin{equation}
z(y,\dot{y})=\alpha\dot{y}-\beta(\gamma|\dot{y}||z|^{\nu-1}+\delta\dot{y}|z|^{\nu})\label{2},
\end{equation}
where the Bouc-Wen parameters $\alpha$, $\beta$ , $\gamma$ ,$\delta$ and $\nu$ are used to tune the shape and smoothness of the system hysteresis loop.

The data are generated by integrating \eqref{1} and \eqref{2} using Newmark integration at a sampling rate of 15,000Hz, and then low-pass filtered and down sampled to 750Hz. There are totally 40,960 training data points and two validation sets, respectively 8192 multi-sine input and 153000 swept sine input. According to \cite{noel2016hysteretic}, we use the simulated RMSE to evaluate the performance of system, and report it as the dB form, i.e., $20\lg(\mathrm{RMSE})$, in which RMSE stands for the root mean square error and can be expressed as,
\begin{equation}
\mathrm{RMSE}=\sqrt{\frac{1}{N_s}\sum_{t=1}^{N}(y(t)-{y}_s(t))^2},
 \end{equation}
 in which $N_s$ is the number of validation data points. 

The same as  \cite{westwick2018using}, we choose $30$ regressors and all of the 40960 samples are used for training. Then the initial value of the linear transformation matrix is identified according to Section 4.1, i.e., the FROLS method was used to establish an NARX polynomial model and obtain the Hessian, then the Hessian at all samples are stacked and the decoupling method is used to obtain the initial $\mathbf V$. We set  $n=5$ and $q=10$, then $\mathbf V\in \mathbb{R}^{30\times5}$, $\mathbf X\in \mathbb{R}^{40960\times 5}$, $\mathbf B\in \mathbb{R}^{40960\times 51}$ and $\mathbf w\in \mathbb{R}^{51}$. After the initialization, the Variable Projection Method was used to estimate all the parameters and the number of iterations is 43. 

It is noted that the condition numbers of $\mathbf{U}$ and $\mathbf{X}$ are $1.74 \times 10^7$ and 28.26, respectively. In fact, the regressors in $\mathbb{U}$ are highly correlated, as the system inputs and outputs at different time instants are actually dependent on each other. After the linear transformation, uncorrelated variables are formed, making the following identification easier, confirming the effectiveness of the dimensionality reduction.  For the 5 selected variables, the variable space is partitioned into $10^5$ subregions, and in each subregion, the approximated nonlinear dynamic system is linear.

\begin{figure}
	\begin{center}
		\includegraphics[width=0.8\columnwidth]{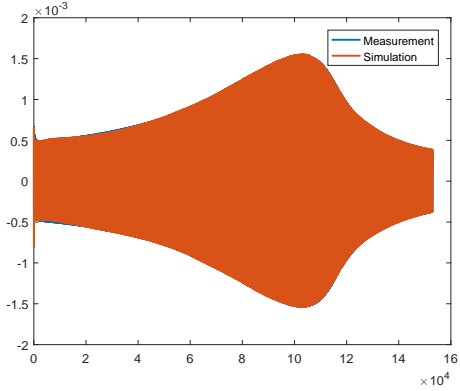}    
		\caption{The simulated and measured output on the swept sine validation Data.} 
		\label{fig:swe}
	\end{center}
\end{figure}

\begin{figure}
	\begin{center}
		\includegraphics[width=0.8\columnwidth]{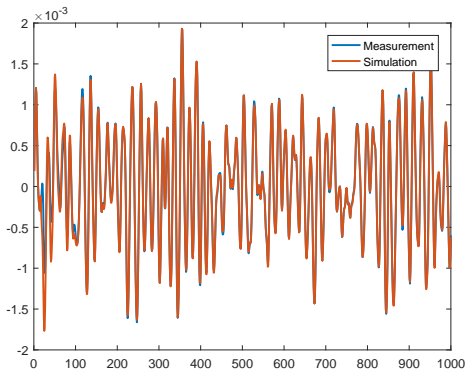}    
		\caption{The simulated and measured output on the multi Sine validation Data.} 
		\label{fig:mul}
	\end{center}
\end{figure}

Table \ref{tb:bouc} lists the simulation error when using the UReLU neural network, with both the multi-sine and swept sine inputs, denoted by RMSE(mul) and RMSE(swe), respectively. The results are compared with other state-of-the-art results as well as those obtained with a single layer ReLU neural network which contains 50 neurons. The number of parameters used for each method are also listed in Table \ref{tb:bouc}, in which the notation ``-'' indicates that the corresponding value is not available.  Fig \ref{fig:swe} and \ref{fig:mul} show the simulated outputs of the UReLU approximated system evaluated on the 2 validation data sets, and the measured output  are also plotted for comparison.
\begin{table*}[htbp]
	\begin{center}
		\caption{Comparison of the test performance of several approaches on Bouc-Wen system.}\label{tb:bouc}
		\begin{tabular}{cccc}
			\toprule[1.5pt]  
			  Method & RMSE(mul) & RMSE(swe) &\# Parameter\\
			 \midrule[1.2pt]
						 NARX (\cite{belz2017automatic}) & -75.73 & -77.20 &-\\
			 \midrule
			EHH   (\cite{xu2019efficient})& -83.00 & -88.78& 3530\\ 
			 \midrule
			 D-NARX (\cite{westwick2018using}) & -85.42 & -95.55 &206\\
			\midrule
			ReLU & -74.64 & -73.99&1550 \\
			 \midrule
			 UReLU & -87.18 & -96.41&201 \\
			\bottomrule[1.5pt] 
		\end{tabular}
	\end{center}
\end{table*}


It can be seen that the result of UReLU neural network is quite encouraging. Compared with other method, we have achieved the best accuracy. Besides, the number of parameters employed in the UReLU neural network is also the smallest, i.e., there are only 201 parameters in the UReLU network while in the single layer ReLU neural network with the same number of neurons, the number of parameters is 1550.


\section{Conclusions and future work}


In this paper, the UReLU neural network is proposed based on the linear transformation and UReLU function, which can be seen as the decoupled ReLU neural network and  the interpretability can be shown clearly.  The training of the UReLU neural network follows a decoupled strategy, and 
the easy obtained derivative of the UReLU neural network facilitates the training process. 
In the simulation study,  the UReLU neural network is used to approximate a complex nonlinear system and the performance is excellent.  

Currently, we consider only the single hidden layer neural network, and the next step is to increase the number of layers of the network to get better accuracy when dealing with large-scale problems and at the same time retain interpretability.


\begin{thebibliography}{25}
\providecommand{\natexlab}[1]{#1}
\providecommand{\url}[1]{\texttt{#1}}
\providecommand{\urlprefix}{URL }
\expandafter\ifx\csname urlstyle\endcsname\relax
  \providecommand{\doi}[1]{doi:\discretionary{}{}{}#1}\else
  \providecommand{\doi}{doi:\discretionary{}{}{}\begingroup
  \urlstyle{rm}\Url}\fi

\bibitem[{Abdalmoaty and Hjalmarsson(2018)}]{abdalmoaty2018application}
Abdalmoaty, M.R. and Hjalmarsson, H. (2018).
\newblock Application of a linear {PEM} estimator to a stochastic
  wiener-hammerstein benchmark problem.
\newblock \emph{IFAC-PapersOnLine}, 51(15), 784--789.

\bibitem[{Belz et~al.(2017)Belz, M{\"u}nker, Heinz, Kampmann, and
  Nelles}]{belz2017automatic}
Belz, J., M{\"u}nker, T., Heinz, T.O., Kampmann, G., and Nelles, O. (2017).
\newblock Automatic modeling with local model networks for benchmark processes.
\newblock \emph{IFAC-PapersOnLine}, 50(1), 470--475.

\bibitem[{Billings(2013{\natexlab{a}})}]{Billings2013}
Billings, S. (2013{\natexlab{a}}).
\newblock \emph{{Nonlinear System Identification: NARMAX Methods in the Time,
  Frequency and Spatio-Temporal Domains}}.
\newblock Wiley.

\bibitem[{Billings et~al.(1988)Billings, Korenberg, and
  Chen}]{billings1988identification}
Billings, S., Korenberg, M., and Chen, S. (1988).
\newblock Identification of non-linear output-affine systems using an
  orthogonal least-squares algorithm.
\newblock \emph{International Journal of Systems Science}, 19(8), 1559--1568.

\bibitem[{Billings(2013{\natexlab{b}})}]{billings2013nonlinear}
Billings, S.A. (2013{\natexlab{b}}).
\newblock \emph{Nonlinear system identification: NARMAX methods in the time,
  frequency, and spatio-temporal domains}.
\newblock John Wiley \& Sons.

\bibitem[{Billings and Wei(2005)}]{billings2005new}
Billings, S.A. and Wei, H.L. (2005).
\newblock A new class of wavelet networks for nonlinear system identification.
\newblock \emph{IEEE Transactions on neural networks}, 16(4), 862--874.

\bibitem[{Chen and Billings(1992)}]{chen1992neural}
Chen, S. and Billings, S. (1992).
\newblock Neural networks for nonlinear dynamic system modelling and
  identification.
\newblock \emph{International Journal of control}, 56(2), 319--346.

\bibitem[{Dreesen et~al.(2018)Dreesen, De~Geeter, and
  Ishteva}]{dreesen2018decoupling}
Dreesen, P., De~Geeter, J., and Ishteva, M. (2018).
\newblock Decoupling multivariate functions using second-order information and
  tensors.
\newblock In \emph{International Conference on Latent Variable Analysis and
  Signal Separation}, 79--88. Springer.

\bibitem[{Dreesen et~al.(2017)Dreesen, Westwick, Schoukens, and
  Ishteva}]{dreesen2017modeling}
Dreesen, P., Westwick, D.T., Schoukens, J., and Ishteva, M. (2017).
\newblock Modeling parallel wiener-hammerstein systems using tensor
  decomposition of volterra kernels.
\newblock In \emph{International Conference on Latent Variable Analysis and
  Signal Separation}, 16--25. Springer.

\bibitem[{Ganjefar and Tofighi(2015)}]{ganjefar2015single}
Ganjefar, S. and Tofighi, M. (2015).
\newblock Single-hidden-layer fuzzy recurrent wavelet neural network:
  Applications to function approximation and system identification.
\newblock \emph{Information Sciences}, 294, 269--285.

\bibitem[{Golub and Pereyra(2003)}]{golub2003separable}
Golub, G. and Pereyra, V. (2003).
\newblock Separable nonlinear least squares: the variable projection method and
  its applications.
\newblock \emph{Inverse problems}, 19(2), R1.

\bibitem[{Hochreiter(1998)}]{hochreiter1998vanishing}
Hochreiter, S. (1998).
\newblock The vanishing gradient problem during learning recurrent neural nets
  and problem solutions.
\newblock \emph{International Journal of Uncertainty, Fuzziness and
  Knowledge-Based Systems}, 6(02), 107--116.

\bibitem[{Kolda and Bader(2009)}]{kolda2009tensor}
Kolda, T.G. and Bader, B.W. (2009).
\newblock Tensor decompositions and applications.
\newblock \emph{SIAM review}, 51(3), 455--500.

\bibitem[{Krizhevsky et~al.(2012)Krizhevsky, Sutskever, and
  Hinton}]{krizhevsky2012imagenet}
Krizhevsky, A., Sutskever, I., and Hinton, G.E. (2012).
\newblock Imagenet classification with deep convolutional neural networks.
\newblock In \emph{Advances in neural information processing systems},
  1097--1105.

\bibitem[{Lathi(1998)}]{lathi1998signal}
Lathi, B.P. (1998).
\newblock \emph{Signal processing and linear systems}.
\newblock Oxford University Press New York.

\bibitem[{Lawton and Sylvestre(1971)}]{Lawton1971Elimination}
Lawton, W. and Sylvestre, E. (1971).
\newblock Elimination of linear parameters in nonlinear regression.
\newblock \emph{Technometrics}, 13(3), 7.

\bibitem[{Lennart(1999)}]{lennart1999system}
Lennart, L. (1999).
\newblock System identification: theory for the user.
\newblock \emph{PTR Prentice Hall, Upper Saddle River, NJ}, 1--14.

\bibitem[{Nelles(2013)}]{nelles2013nonlinear}
Nelles, O. (2013).
\newblock \emph{Nonlinear system identification: from classical approaches to
  neural networks and fuzzy models}.
\newblock Springer Science \& Business Media.

\bibitem[{Neyshabur et~al.(2016)Neyshabur, Wu, Salakhutdinov, and
  Srebro}]{neyshabur2016path}
Neyshabur, B., Wu, Y., Salakhutdinov, R.R., and Srebro, N. (2016).
\newblock Path-normalized optimization of recurrent neural networks with relu
  activations.
\newblock In \emph{Advances in Neural Information Processing Systems},
  3477--3485.

\bibitem[{No{\"e}l and Schoukens(2016)}]{noel2016hysteretic}
No{\"e}l, J. and Schoukens, M. (2016).
\newblock Hysteretic benchmark with a dynamic nonlinearity.
\newblock In \emph{Workshop on nonlinear system identification benchmarks},
  7--14.

\bibitem[{Schoukens et~al.(2016)Schoukens, Vaes, and
  Pintelon}]{schoukens2016linear}
Schoukens, J., Vaes, M., and Pintelon, R. (2016).
\newblock Linear system identification in a nonlinear setting: Nonparametric
  analysis of the nonlinear distortions and their impact on the best linear
  approximation.
\newblock \emph{IEEE Control Systems Magazine}, 36(3), 38--69.

\bibitem[{Vervliet et~al.(2016)Vervliet, Debals, Sorber, Van~Barel, and
  De~Lathauwer}]{vervliet2016tensorlab}
Vervliet, N., Debals, O., Sorber, L., Van~Barel, M., and De~Lathauwer, L.
  (2016).
\newblock Tensorlab 3.0. available online.
\newblock \emph{URL: http://www. tensorlab. net}.

\bibitem[{Westwick et~al.(2018)Westwick, Hollander, Karami, and
  Schoukens}]{westwick2018using}
Westwick, D.T., Hollander, G., Karami, K., and Schoukens, J. (2018).
\newblock Using decoupling methods to reduce polynomial narx models.
\newblock \emph{IFAC-PapersOnLine}, 51(15), 796--801.

\bibitem[{Xu et~al.(2019)Xu, Tao, Li, Xi, Suykens, and Wang}]{xu2019efficient}
Xu, J., Tao, Q., Li, Z., Xi, X., Suykens, J.A., and Wang, S. (2019).
\newblock Efficient hinging hyperplanes neural network and its application in
  nonlinear system identification.
\newblock \emph{arXiv preprint arXiv:1905.06518}.

\bibitem[{Yao et~al.(2018)Yao, Wang, and Zhang}]{yao2018identification}
Yao, X., Wang, Z., and Zhang, H. (2018).
\newblock Identification method for a class of periodic discrete-time dynamic
  nonlinear systems based on sinusoidal esn.
\newblock \emph{Neurocomputing}, 275, 1511--1521.

\end{thebibliography}
\end{document}